\documentstyle[epsf,referee,12pt]{l-aa}
\def\beginrefs{\begingroup\parindent=0pt\frenchspacing
   \parskip=1pt plus 1pt minus 1pt\interlinepenalty=1000\tolerance=400
   \everypar={\hangindent=0.42in}%\hyphenpenalty=10000
   }
\def\endrefs{\endgroup\vfill\eject}

\def \ergs {ergs~s$^{-1}$~cm$^{-2}$}

\begin{document}
%\doublespace
   \thesaurus{(12.04.2;  %Cosmology: diffuse radiation
               13.25.3;  %X-rays: general
               11.17.3)} %quasars: general
   \title{QSOs AND THE HARD X-RAY BACKGROUND}

   \author{A. Vikhlinin
%          \inst{1}
          }

   \offprints{A. Vikhlinin}

   \institute{Space Research Institute, Russian Academy of Sciences\\
              Profsoyuznaya 84/32,Moscow 117810, Russia\\
              internet: vikhlinin@hea.iki.rssi.ru
             }

 \date{\today}
   \maketitle

%\clearpage
\begin{abstract}
%\singlespace
We calculate the contribution to the cosmic x-ray background (CXB) of a
population of power law spectrum sources with spectral indices distributed
over a broad range of values. The composite spectrum of this source
population is significantly harder than that given by the power law having
the average value of spectral indices. Starting from spectral distributions
which are approximately those observed from quasars, it is possible to
reproduce the CXB spectrum from $\sim 0.5$ keV to $\sim 20$ keV. If the
spectra of quasars steepen at around 100 keV, the resulting composite
spectrum nearly perfectly fits the CXB in the even broader energy range, up
to $\sim 100$ keV. The QSO population with broadly distributed spectral
parameters is also characterized by a significant discrepancy between the
results of hard and soft x-ray source counts.  The same population of
sources yields about three times more sources at 10 keV than at the
corresponding flux at 1 keV, similarly to what is found from the comparison
of HEAO A-1/Ginga and Einstein/ROSAT measurements. Thus, by allowing the
spectra of QSO's to span a broad range of spectral indices, it is possible
to reproduce both the CXB spectrum and account for the apparent differences
in number counts in different energy bands.


\end{abstract}

%\singlespace
%\clearpage

\section{Introduction} \label{sec-intro}
In spite of great progress in the study of the comic x-ray background (CXB)
radiation achieved since its discovery more than 30 years ago (Giacconi et
al. 1962) its origin remains unclear. In the soft x-ray band, deep ROSAT
source counts show that the population of faint x-ray sources (of them some
70\% are QSOs, Boyle et al. 1993) is able to account for the main portion
(if not 100\%) of the soft x-ray (1-2 keV) background, while the
contribution of a truly diffuse component (such as the hot intergalactic
gas) must be less than 25\% at 90\% confidence (Hasinger et al. 1993).

At harder x-rays, 3-60 keV, sensitive imaging surveys have not yet been
performed, and the main efforts have been concentrated on explaining the CXB
spectrum which is remarkably well fit by the thermal bremsstrahlung model
with $kT=40$ keV (Marshall et al. 1980).  The COBE limit on the
Comptonization parameter excludes the possibility of a uniform hot medium
producing more than a few percent of the CXB in the harder x-ray band
(Mather et al. 1990). Hence, most likely the hard x-ray background is also
composed by unresolved compact sources (see the discussion of the ``spectral
coincidence'' and the contribution of point sources to the CXB in Giacconi
\& Zamorani 1987).

The main difficulty with the discrete source explanation of the hard x-ray
background has been that there is no known source population whose spectra
resemble that of the CXB. Clusters of galaxies, which are as numerous as
AGNs at high fluxes in the standard x-ray band (Piccinotti et al. 1982),
typically have thermal bremsstrahlung spectra with temperatures of about 6
keV (David et al. 1993), and exhibit negative cosmological evolution (e.g.,
Gioia et al. 1990).  Hence, they cannot be significant contributors to the
broad-band x-ray background.  Population of AGN's which dominate the soft
x-ray source counts and exhibit positive cosmological evolution have 2--20
keV spectra significantly steeper than the x-ray background. HEAO
A-1/EXOSAT/Ginga observations of low luminosity AGNs (mostly Seyfert 1
galaxies) revealed a ``canonical'' $\alpha=0.7$ spectrum (Mushotzky 1984),
to be compared with the energy index of 0.4 for the CXB in this energy band.
Both soft and hard x-ray observations of high luminosity AGNs (mostly
quasars) suggested that there was no universal power law for this class of
sources.  Observed spectral indices are distributed from $\alpha \sim 0$ to
$\alpha \sim 2.0$, with the mean value of about 1 in the soft x-rays and
0.8-0.9 in the 2-10 keV x-ray band (Williams et al. 1992, Comastri et al.
1992, and Wilkes \& Elvis 1987).  The average spectral index of quasars also
corresponds to a spectrum which is significantly softer than that of the
CXB, which was used to argue that their contribution to the hard x-ray
background could not exceed $\sim 10$\% (Fabian, Canizares, \& Barcons
1989).

In this paper we show that it is still possible to make a significant
portion of the hard x-ray background from QSOs, i.e. that the observed
composite spectrum of quasars can be as hard as the CXB spectrum.  The basic
idea relies on the observed broad spread of QSO spectral indices. X-rays
observed from QSOs at energy $E$ are emitted at higher energies, $(1+z)E$.
Therefore, distant objects with flat spectra appear brighter than those with
steep spectra, and hence, it follows that the mean spectrum hardens with
increasing redshift even though no true spectral evolution occurs. A similar
effect arises if soft x-ray luminosities do not correlate with spectral
properties.  The absence of such a correlation implies that hard spectrum
sources should be on average more luminous in the 2-10 keV x-ray band than
the steep spectrum sources. Futhermore, the broad distribution of spectral
indices provides a natural explanation for the mismatch between the results
of soft and hard x-ray source counts found from the comparison of
Einstein/ROSAT and HEAO A-1/Ginga surveys.  Since hard sources are brighter
at high redshifts than soft ones, a difference in number-flux relations
determined at, for example, 1 keV and 10 keV is expected, with larger
surface number density at the hard energy band.

In section \ref{sec-chatter} we argue that it is necessary to account for
the contribution of QSOs in source models of the x-ray background. Models of
the QSO population (spectra, luminosity function, and the cosmological
evolution) used in the paper are described in section \ref{sec-models}.
Basic equations used for calculations of the composite QSO spectrum and the
number-flux relations in different energy bands are derived in section
\ref{sec-formulas}.  Section \ref{sec-results} presents the results of
calculations. The main results of the paper are briefly discussed in section
\ref{sec-discussion}.

\section {QSOs, Seyfert Is, and the CXB} \label{sec-chatter}

The difficulties with the explanation of the CXB spectrum by AGNs arise from
the assumption that their spectra are simple power laws over a broad energy
range. Schwartz \& Tucker (1988) have shown that if AGN spectra flatten with
increasing energy up to some energy cutoff, these difficulties can be
overcome and it is possible to match the CXB spectrum by the composite
spectrum of unresolved distant AGNs. In fact, the spectral flattening was
observed in Ginga spectra of Seyfert 1 galaxies and was interpreted as
either the partial covering by cold material of a power law x-ray spectrum
or as reprocessed emission from cold material (Piro et al. 1989, Matsuoka et
al. 1990, Pounds et al. 1990). GRANAT and OSSE observations of a sample of
Seyfert 1s revealed that their spectra do steepen again at energies of about
50-100 keV (Jourdain et al. 1992, Johnson et al.  1994).  Thus the observed
spectra of Seyfert 1s are quite close to what is required by the model of
Schwartz \& Tucker (1988) -- the source spectra flatten above $\sim 10$ keV
and they steepen again at about 100 keV. This stimulated Zdziarski et al.
(1993) to claim that it is AGNs with Seyfert~1--like spectra which comprise
the 2--100 keV x-ray background. In fact, using the cosmological evolution
model of Boyle et al. (1993) and the model spectrum determined from
Ginga/OSSE observations of Seyfert 1s Zdziarski et al. (1993) obtained an
excellent fit to the CXB spectrum in the 2-100 keV energy range.

However, the explanation of the CXB by Seyfert 1 galaxies may fail because,
as discussed by Giacconi \& Zamorani (1987), the contribution of known
classes of steep spectrum sources must also be included in modeling the CXB
spectrum such as quasars and clusters of galaxies. Giacconi \& Zamorani
showed that if the contribution of the $\alpha \sim 0.7-1$ sources to the
soft x-ray background amounts to at least 50\% of the CXB, the residual CXB
spectrum is much flatter than the $kT=40$ keV spectrum.  Therefore, it may
be difficult to fit the residual CXB spectrum by spectra of Seyfert 1
galaxies.

Deep, soft surveys show that the contribution of QSO's to the CXB is in fact
significant. ROSAT deep surveys show that point-like faint sources
contribute the bulk of the 2 keV CXB, with the directly resolved fraction
being about 60\%. From fluctuation analyses, point sources contribute about
75\% of the CXB and a reasonable extrapolation of the number-flux relation
to even fainter fluxes can account for 100\% of the soft x-ray background
(Hasinger et al.  1993). In a deep ROSAT survey, Boyle et al.  (1993)
identify 64\% of the extragalactic x-ray sources with QSOs and this number
can be as high as $\sim 80$\%. Hence, it seems very likely that QSOs
contribute not less 50\% of the 2 keV CXB, and consequently their
contribution cannot be overlooked in any model of the CXB.

QSO spectra are generally less well studied than those of Seyfert 1
galaxies.  Both soft and hard x-ray observations show that spectral indices
of quasars are distributed over a broad range of values, from $\alpha \sim
0$ to $\alpha \sim 2$, and the mean spectral index in the 2-10 keV band is
0.8-0.9 (Comastri et al. 1992, Williams et al. 1992). In their review of
x-ray observations of AGNs, Mushotzky, Done \& Pounds (1993) discuss the
difference between 2-10 keV spectra of Seyfert 1 galaxies and quasars.
Ginga spectra show that the radio-loud QSOs do not require additional
reflection components and, apart from 3C273 and PHL 1657, show no deviations
from a simple power law over the 2--20 keV energy band (Williams et al.
1992). The radio-quiet QSO spectra do show deviations from the power law but
the steeper underlying spectrum ($\alpha \approx 1.0$) implies a real
difference between radio-quiet QSOs and Seyfert 1 galaxies. We conclude
that QSOs are different from Seyfert Is from the point of view of x-ray
spectra and show less evidence for spectral flattening above 10 keV.
Therefore, it is necessary to subtract the contribution of QSOs to the x-ray
background before fitting the CXB spectrum by spectra of Seyfert Is. If we
assume that QSOs all have the $\alpha=0.8-0.9$ spectrum and contribute about
50\% to the 2 keV background, the spectrum of the residual background will
be much harder than the $kT=40$ keV spectrum (Giacconi \& Zamorani 1987),
which will be difficult to fit even with spectra of Seyfert Is with
significant reflection contributions.

\section{Models of The Source Population} \label{sec-models}

We consider four models of the source population to reconstruct the CXB
spectrum and to illustrate properties of the reconstruction -- ``realistic''
(model {\bf A}), ``reduced realistic'' ({\bf B}), ``pure'' ({\bf C}), and
``common'' ({\bf D}). Model spectra are single power laws modified at the
soft and the hard energies to describe a soft excess and a high energy
cutoff:
\begin{eqnarray}
F(E) & = & \left\{\begin{array}{ll}
		a_sE^{-\alpha_s} & \mbox{$E<E_s$}\\
		a_mE^{-\alpha_m} & \mbox{$E_s<E<E_h$}\\
		a_hE^{-\alpha_h} & \mbox{$E>E_h$}
                                   \end{array}
                                        \right.
\end{eqnarray}

In the standard x-ray band (2-10 keV) we assume that sources have power law
spectra, with no additional features such as strong emission lines,
absorption edges, or a reflection component, and have spectral indices
distributed according to a Gaussian distribution with the mean value of 0.8
and the dispersion of 0.35. ``Unphysical'' energy spectra with negative
spectral indices are prohibited in the model by a cutoff of the distribution
at zero.  The choice of parameters is based on the results of {\em EXOSAT}\/
and {\em Ginga}\/ measurements of spectra of subsamples of soft x-ray
selected high luminosity AGN (Comastri et al. 1992, Williams et al.  1992).
Both sets exhibit a steep 2-10 keV mean spectral index (0.81 for {\em
Ginga}\/ and 0.89 for {\em EXOSAT}\/ data), with a wide range of spectral
indices, distributed from 0.08 to 1.26, with an intrinsic dispersion of
about 0.3.  There is evidence of a soft excess in QSO energy spectra below
an energy of a few keV (Wilkes \& Elvis 1987, Comastri et al.  1992, Fiore
et al.  1994), manifesting itself as a gradual steepening of source spectra
in softer energy bands.  Hence, we introduce the soft excess to the model
spectra as an additional power law below 1 keV, with the mean soft x-ray
spectral index of 1.4. The choice of a mean soft x-ray spectral index is
motivated by ROSAT measurements of bright QSO spectra in the energy band of
0.1-2.4 keV (Laor et al. 1994).

The bulk of sources in the model have spectral indices less than 1, which
yields infinite broad band luminosities. Therefore, the break in spectra at
high energies is required. We introduce such a break at around 100 keV by
steepening the spectral index up to the value of 2.0.  The cutoff energy of
100 keV is chosen so that it is possible to reproduce the CXB spectrum up to
$\sim 100$ keV, rather that this choice is motivated by the observations of
QSO spectra.

Parameters of the soft excess and the high energy cutoff are distributed
according to a Gaussian distribution with a dispersion equal to 0.3 of the
mean value. For the high energy cutoff, the extremely low values of the
break energy (below 30 keV) and flat spectra (with the hard spectral index
less than the main spectral index) are prohibited.

The four models differ by the distribution of spectral parameters. In the
``realistic'' model {\bf A} all the spectral parameters are characterized by
Gaussians.  The ``reduced realistic'' model {\bf B} is used to show how the
range in the soft excess and the high energy cutoff parameters influence the
composite spectrum; the soft excess parameters are held fixed at a value of
1.4 for the spectral index and 1 keV for the break energy, and the high
energy cutoff parameters at values of 2.0 for the spectral index and 100 keV
for the break energy. The ``pure'' model {\bf C} is used to demonstrate the
effects of the soft excess and the high energy cutoff and is characterized
by single power law spectra over the whole energy range with a distribution
of spectral indices analagous to that in the model {\bf A}.  Finally, in the
``common'' model {\bf D} all parameters are held fixed which corresponds to
the assumption of an identical spectrum for all sources.

We assume further that there is no correlation between the spectral
properties and the luminosity in the soft x-ray band. This would of course
imply the existence of a negative correlation between the luminosity in the
harder x-ray band (2-10 keV) and the spectral index, i.e.  hard sources
must be on average more luminous in this band. Existing x-ray data do not
enable one to explore the existence of such a correlation with high
confidence, since the 2-10 keV x-ray spectra are measured only for the
several dozens of the brightest quasars. However, Williams et al. (1992),
found a negative correlation between the x-ray luminosity and 2-10 keV
spectral index at more than 90\% significance in their sample of 13
high-luminosity AGNs, while Wilkes \& Elvis (1987) find no correlation
between the x-ray luminosities and spectral indices measured in the 0.3-3.5
keV band. Finally, recent ROSAT measurements indicate that quasar intensities
are much less scattered at around 0.3 keV than at higher energies (Laor et
al. 1994).

For the model luminosity function we use the QSO soft x-ray luminosity
function derived by Boyle et al. (1993) from the joint analysis of the deep
ROSAT survey, and the EMSS luminosity function. At $z=0$, this luminosity
function can be represented as two power laws, with a break at
$L^*_X=10^{43.9}$ erg s$^{-1}$ in the 0.3-3.5 keV energy band,
$\Phi(L_X)\propto L_X^{-3.4}$ above the break, and $\Phi(L_X)\propto
L_X^{-1.7}$ below the break. Boyle et al. (1993) also derive the
cosmological evolution of QSOs which can be described as  pure luminosity
evolution, $L_X \propto (1+z)^K$, with $K=2.8$ for $\Omega=0$ and $K=2.5$
for $\Omega=1$ universe (Boyle et al. 1993). For $\Omega=1$ a ``cut off''
in the  evolution is required at around $z=2$. We use the same model of
cosmological evolution for the values of $\Omega$ ranging from $\Omega=0$
and $\Omega=1$ in our calculations, with $K=2.7$, and cutoff of evolution at
$z=3.0$.

It must be emphasized that no spectral evolution is present in calculations.
This means that, by assumption, the distribution of spectral parameters is
the same at every redshift.

\section{Basic Equations} \label{sec-formulas}

If at the source rest-frame the source  spectrum is given as $A\,S_0(E)$, and
the source lies
at a redshift $z$, then the observed energy flux density (flux per unit
energy interval) is:
\begin{eqnarray}	\label{eq-flux}
f(E) & = & \frac{A\,S_0(E(1+z))}{4\pi(a_0 r)^2 (1+z)},
\end{eqnarray}
where $r(z,\Omega)$ is the angular size distance,
\begin{equation}	\label{eq-angsize}
a_0\, r(z,\Omega)\, (H_0/c) = \frac{2}{\Omega^2(1+z)}\;\{\Omega z +
(\Omega-2)\,[\sqrt{1+\Omega z} -1]\}
%\begin{eqnarray}	\label{eq-psi}
%\Psi(z,\Omega) & = & \frac{2}{\Omega^2(1+z)^2}\;\{\Omega z +
%(\Omega-2)\,[\sqrt{1+\Omega z} -1]\}
\end{equation}
(Mattig 1958). For power law spectra,
$S_0(E)=E^{-\alpha}$, eq. \ref{eq-flux} becomes:
\begin{eqnarray}
f(E) & = & \frac{A\,E^{-\alpha}\,(1+z)^{-\alpha-1}}{4\pi(a_0r)^2}.
\end{eqnarray}
% from which it follows that sources with flat spectra (i.e. with smaller
% $\alpha$) are apparently brighter at high redshifts than sources with steep
% spectra. If at zero redshift the luminosities of soft and hard sources are
% approximately the same, at high redshift hard sources appear brighter than
% soft ones. This effect gives rise to an apparent ``hardening'' of the
% composite spectrum with increasing redshift and is the key to generating
% the CXB spectrum from a source population exhibiting no spectral evolution
% and whose mean spectral distribution, at the present epoch, is softer than
% the CXB spectrum.

If sources are distributed homogenously in space and there is no density
evolution, the number of sources per unit redshift increment is
\begin{eqnarray}	\label{eq-dn}
\frac{dN}{dz} & = & n_0 \;\frac{4\pi c^3}{H_0^3}\; z^2 \frac {4[\Omega z +
(\Omega-2)(\sqrt{1+\Omega z}-1)]^2} {\Omega^4 z^2 (1+z)^3\sqrt{1+\Omega z}}
\;\; =\;\; n_0 \;\frac{4\pi c^3}{H_0^3}\; z^2\; \xi(z,\Omega)
\end{eqnarray}
where $n_0$ is the volume number density at z=0 (Mattig 1958).

Einstein and ROSAT observations indicate that the observed properties of the
high luminosity AGN population require a cosmological evolution which can
be described in terms of pure luminosity evolution, i.e. the luminosity
scales with redshift as $L_x(z)=L_x(0)\,C(z)$, with $C(z)=(1+z)^K$ (Boyle et
al. 1993).  For this type of evolution the $z$-dependence of the luminosity
function is:
\begin{eqnarray}	\label{eq-lfunctionz}
\Phi(L_x,z) & = & \frac{1}{C(z)}\;\Phi\left(\frac{L_x}{C(z)},0\right)
\end{eqnarray}
Available x-ray data on the luminosity function and its evolution is derived
from observations in the 0.3-3.5 keV energy band. For simplicity, we assume
that the the same relations apply to 1 keV luminosities; $L_1$ is denoted as
the 1 keV luminosity in the source rest-frame.

Using equations \ref{eq-flux}-\ref{eq-lfunctionz} we calculate the composite
energy spectrum of the source population, and the results of source counts
as a function of energy band.
% Our calculations assume that the spectral
% properties do not correlate with redshift or with the x-ray luminosity at
% around 1 keV (justifications for these assumptions are given in section
% \ref{sec-models}).
The spectral parameter $\alpha$ is used to
designate different spectral shapes. For a class of power-law spectra
$\alpha$ is simply the power-law index; for more complicated spectral shapes
$\alpha$ becomes a vector of parameters, and integrals over the range of
$\alpha$ indicate integrals over different spectral shapes. The assumed
absence of any correlation of spectral properties with redshift or with
soft x-ray luminosity (section \ref{sec-models}) means that the density
distribution of the spectral
parameter, $\alpha$, is a function of $\alpha$ only, i.e.  the probability
that $\alpha$ is within the interval $(\alpha,\alpha+d\alpha)$ is
\begin{eqnarray}
dP & = & W(\alpha)\,d\alpha
\end{eqnarray}
Different spectral shapes are denoted as $S_0(E,\alpha)$, with this function
normalized to unity at 1 keV, so that the energy flux density at a
particular energy $E_0$ is given by $L_1 S_0(E_0,\alpha)$, where $L_1$ is
the 1~keV source luminosity.

\subsection {Composite Spectrum}

In this section we calculate the composite spectrum of sources uniformly
distributed in space, with spectral properties described above, and pure
luminosity evolution.
% As was already discussed, if the soft x-ray luminosity
% does not correlate with the spectral shape, hard sources should appear
% more luminous at higher energies. As a result, the composite spectrum at
% high energies is harder than that given by the mean values of the spectral
% parameters. For example, in the case of power law spectra, the composite
% spectrum is harder at high energies than the power law with the mean
% spectral index.
Let us first find the mean energy spectrum of sources located at a redshift
$z$. This can be evaluated by integration of equation \ref{eq-flux} over the
appropriate range of luminosities and spectral shapes:
\begin{eqnarray}	\label{eq-cxbspec-z1}
F_{z}(E) & = &  \int\limits_0^\infty \Phi(L_1,z)\,dL_1\;\; L_1
\int\limits_{\alpha}W(\alpha)\,d\alpha\;
\frac{S_0(E(1+z),\alpha)}{4\pi\,(a_0r)^2\;(1+z)}
\end{eqnarray}
For pure luminosity evolution the luminosity function changes with redshift
following eq.\ref{eq-lfunctionz}, and the evaluation of the integral over
$L_1$ in eq.\ref{eq-cxbspec-z1} yields:
\begin{eqnarray}	\label{eq-cxbspec-z}
F_{z}(E) & = &  \overline{L}_1\;\frac{C(z)}
{4\pi\,(a_0r)^2\;(1+z)}\;
\int\limits_{\alpha}W(\alpha)\,d\alpha\;
S_0(E(1+z),\alpha)
\end{eqnarray}
where $\overline{L}_1$ is the mean present day soft x-ray luminosity and
$C(z)$ is the cosmological luminosity evolution function.

It is easy to show that equation \ref{eq-cxbspec-z} implies that the
composite spectrum of sources at a given $z$ hardens with increasing
redshift.  Let the source spectra be described as a simple power law,
i.e.  $S_0(E,\alpha)=E^{-\alpha}$. Then equation \ref{eq-cxbspec-z} becomes
\begin{eqnarray}
F_{z}(E) & \propto & \int\limits_{\alpha} W(\alpha)\,d\alpha\, E^{-\alpha}
\,(1+z)^{-\alpha}
\end{eqnarray}
The spectral index of the composite spectrum (as a function of energy) is
the logarithmic derivative of the spectral flux:
\begin{eqnarray}
\overline{\alpha} & = & - \frac {d\,\log F_z(E)}{d\,\log E}
 = - \frac{E}{F_z(E)}\;\frac{d\,F_z(E)}{d\,E} \nonumber \\
	& = & \frac
{
\int\limits_{\alpha} \alpha\,W(\alpha)\,E^{-\alpha}\,(1+z)^{-\alpha}\,d\alpha
}
{
\int\limits_{\alpha} W(\alpha)\,E^{-\alpha}\,(1+z)^{-\alpha}\,d\alpha
},
\end{eqnarray}
i.e. the initial distribution of spectral indices, $W(\alpha)$, must be
replaced by $W(\alpha)\,E^{-\alpha}\,(1+z)^{-\alpha}$ when one observes
sources at a redshift $z$ and an energy $E$. Thus, the distribution is
weighted towards flatter spectra at higher energies and redshifts.

On multiplying eq.\ref{eq-cxbspec-z} by $dN$, the number of sources in the
redshift interval $(z,z+dz)$ (eq. \ref{eq-dn}), and integrating over
redshift, we find the expression for the composite spectrum of the source
population:
\begin{eqnarray} 	\label{eq-cxbspec}
F(E) & = & \left(\frac{c}{H_0}\right)^3\; n_0\, \overline{L}_1 \;
\int\limits_0^\infty dz \;
\frac{\xi(z,\Omega)\; z^2}{(a_0r)^2\;(1+z)}\; C(z) \;
\int\limits_{\alpha} W(\alpha)\,d\alpha\, S_0(E(1+z),\alpha).
\end{eqnarray}
This equation is used in section \ref{sec-results-spec} for calculations of
the composite spectrum of QSOs, using models {\bf A}-{\bf D} for the spectral
parameter distributions.

\subsection {Source counts at different energies}

% As mentioned in the introduction, a broad distribution of spectral
% properties can give rise to an aparent discrepancy between source counts
% performed in hard and the soft x-ray bands. The primary reason for this
% effect is that if the soft x-ray luminosities do not correlate with spectral
% properties, then such a correlation must arise in the hard x-ray band. In
% particular, hard sources must be on average more luminous at higher
% energies (say, 2-10 keV) then sources with steep spectra. Redshift effects
% further enhance the divergence between fluxes of flat- and steep-spectrum
% sources.  Hence, a difference in number-flux relations determined at, for
% example, 1 keV and 10 keV is expected. In this section we derive equations
% which enable one to evaluate the number-flux relation (\lognlogs) as a
% function of energy.

The 1 keV luminosity of a source whose energy flux density observed at the
energy $E_0$ is $f$ can be calculated (cf. eq\ref{eq-flux}) as:
\begin{eqnarray}	\label{eq-L1}
L_1 & = & \frac{4\pi\;(ar_0)^2\;(1+z)}{S_0(E_0(1+z),\alpha)}\; f
\end{eqnarray}
The fraction of sources of a given spectrum, at a given redshift
yielding an observed flux greater than $f$ is calculated by integrating the
luminosity function:
\begin{eqnarray}
n^\prime & = &
\int\limits^\infty_l \Phi(L,z)\,dL
\end{eqnarray}
where the lower limit, $l$ is the limiting luminosity for a given flux,
$f$:
\begin{eqnarray}
l & = & \frac{4\pi\;(a_0r)^2\;(1+z)}{S_0(E_0(1+z),\alpha)}\; f.
\end{eqnarray}
For pure luminosity evolution the redshift dependence of the luminosity
function is given by eq.\ref{eq-lfunctionz}, and the above equation
becomes:
\begin{eqnarray}	\label{eq-n'}
n^\prime & = & \int\limits^\infty_{l/C(z)}
\Phi(L,0)\,dL
\end{eqnarray}
The number-flux relation is obtained by multiplying equation \ref{eq-n'} by
the number of sources in the redshift interval (eq. \ref{eq-dn}) and
integrating over redshift:
\begin{eqnarray}	\label{eq-counts}
N(>f) & = & n_0 \;\frac{4\pi c^3}{H_0^3}\; \int\limits_0^\infty dz \,
\xi (z,\Omega)\; z^2\, C(z) \;
\int\limits_{\alpha} W(\alpha)\, d\alpha
\int\limits^\infty_{l/C(z)}
\Phi(L,0)\,dL
\end{eqnarray}

To compare results of source counts at different energies it is necessary to
express these fluxes in terms of a common energy band. It is common practice
to do this using the mean source spectrum (known or assumed).  Let us denote
$S(E,\alpha_0)$ as the mean source spectrum (i.e.  corresponding to some
mean value of the spectral parameters). The corresponding transformation of
the observed energy flux density $f(E)$ at energy $E$ to flux, $F$, in the
energy band ($E_1,E_2$) is:
\begin{eqnarray}	\label{eq-fluxtransform}
F & = & f(E) \;\;\frac{\int\limits_{E_1}^{E_2} S(E,\alpha_0)\, dE}
{S(E,\alpha_0)}
\end{eqnarray}
Equations \ref{eq-counts} and \ref{eq-fluxtransform} can be used to calculate
difference between the soft and hard x-ray source counts, for example,
at 1 and 10 keV.

\section{Results} \label{sec-results}

In this section we present results of calculations of the CXB spectrum and
the number-flux relation in different energy bands for four models of
spectra distribution (Section \ref{sec-models}) and several values of the
density parameter, $\Omega$.
% Model {\bf A} is the most ``realistic'' one; it assumes that the power law
% spectral indices are distributed according to Gaussian distributions with
% mean value and dispersion close to the values determined from EXOSAT and
% Ginga observations in the 2-10 keV energy band. This model also assumes the
% existence of a soft excess and a high energy cutoff in source spectra.
% Parameters of the soft excess and the high energy cutoff are distributed
% following a Gaussian with a dispersion of 30\% of the mean parameter value.
% In model {\bf B}, the soft excess and the high energy cutoff parameters are
% fixed, to demonstrate how the distribution of these parameters influences
% the results of calculations. In model {\bf C} source spectra contain neither
% a soft excess nor a high energy cutoff. This model illustrates that the main
% results of our study arise from the broad distribution of spectral
% parameters, and are not primarily effected by the existence of additional
% spectral features like the soft excess and the high energy cutoff. Finally,
% calculations of model {\bf D}, in which all the spectral parameters are
% fixed at the mean values, illustrate the results which are obtained under
% the assumption of an identical spectrum for all sources. More detailed
% descriptions of the models used and the justification for the choice of the
% model parameters are given in section
% \ref{sec-models}.
%
No qualitative difference is found in results of calculations performed for
different values of the cosmological density parameter, $\Omega$, ranging
from 0 to 1. When presenting the results we will simply point out the trend
of changing the derived relationships for $\Omega$ increasing from 0 to 1
and will not emphasize the dependence of the results on the density
parameter.

\subsection{Spectrum} \label{sec-results-spec}
\begin{figure}
\epsfysize=6.5in \epsffile{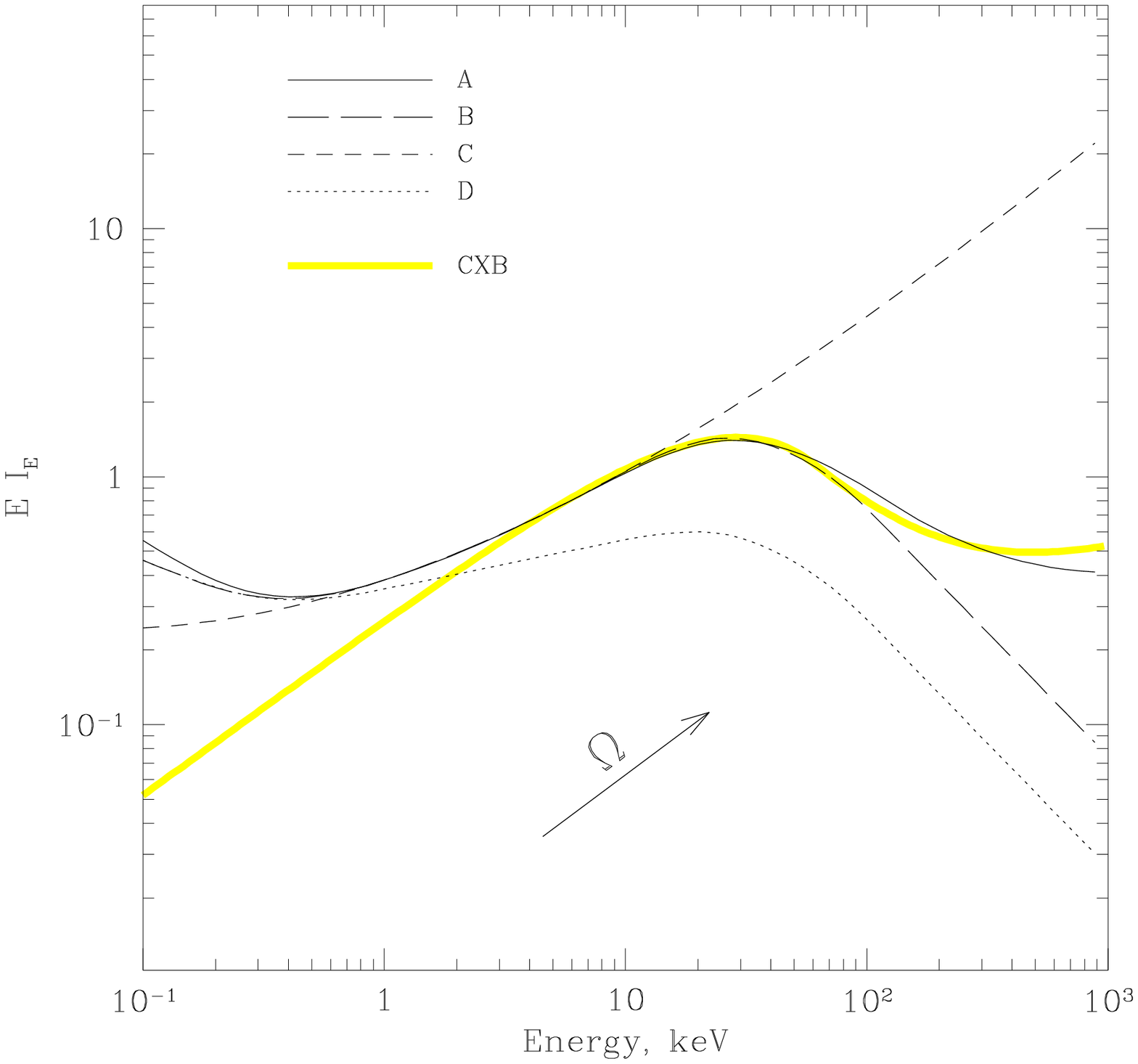}
\caption{
The composite spectrum calculated for models {\bf A}-{\bf D}, and the x-ray
background spectrum. The absolute normalization of model spectra is
arbitrary. All models with the scattering of spectral parameters ({\bf
A}-{\bf C}) adequately describe the CXB spectrum from $\sim 2$ to more than
10 keV.  Models with the high energy cutoff at around 100 keV ({\bf A} and
{\bf B}) succeed in describing the CXB spectrum up to 60-100 keV. Model {\bf
D}, without any scattering of the spectral parameters, reproduces the well
known result: the mean spectrum of QSOs is too steep to fit the CXB spectrum
in the 2-10 keV energy band. The model spectra deviate from the hard x-ray
background extrapolation at energies below 2-2.5 keV. Note that the models
with and without the soft excess in source spectra ({\bf A} \& {\bf B}, and
{\bf C}, respectively) yield almost the same spectra above $\sim 1$ keV.
The arrow shows the trend of model spectra when the density parameter,
$\Omega$, increases.  }

\label{fig-spectrum}
\end{figure}

Composite spectra of the source populations derived for models {\bf A}
through {\bf D} are shown in Fig.\ref{fig-spectrum}. These spectra are
obtained by integrating (eq.\ref{eq-cxbspec}) over redshifts from 0 to 10.
The assumption that sources are still bright at very high redshifts is of
course uncertain. However, we use the model for cosmological evolution in
which the growth of luminosity ceases at a redshift of 3 (section
\ref{sec-models}), and  redshift effects rapidly make the contribution
of sources at higher redshifts negligible (cf.  eq.\ref{eq-cxbspec-z}).

Models with a broad distribution of spectral parameters and the high energy
cutoff ({\bf A} and {\bf B}) perfectly reproduce the CXB spectrum from $\sim
1$ to $\sim 60-100$ keV. Model {\bf C}, without the high energy cutoff,
predicts further flattening of the CXB spectrum above $\sim 20$ keV, but
yields the correct slope of the spectrum in the 2-10 keV energy band. The
``standard'' model {\bf D}, without any scattering of spectral parameters,
reproduces the well-known result -- the average QSO/AGN spectrum is too
steep to fit the CXB spectrum in the energy range from 2-10 keV. Below
$\sim 2$ keV all models predict a steepening of the CXB spectrum, which is
not related to the presence of a soft excess in individual spectra. For
example, model spectrum {\bf C}, without any soft excess lies very close
to spectra {\bf A} and {\bf B} (with soft excess below 1 keV) above 0.7-1
keV.

In the harder x-ray band, above $\sim 20$ keV, a high energy cutoff is
required in source spectra to fit the CXB spectrum. For an $\Omega=0$
Universe, in the context of the present model, the break must be at around
100 keV; this value gradually decreases to 70 keV for $\Omega=1$ (see trend
of model spectra with increasing $\Omega$ indicated by the arrow in Fig.
\ref{fig-spectrum}). The broad distribution of the high energy spectral
index (model {\bf A}) reproduces the flattening of the CXB at $\geq 300$
keV.  However, there is a lack of spectral data on QSO's at such high
energies. Not only are the particular values of the cutoff energies
unknown, but the very existence of breaks in QSO spectra is not very well
established (Mushotzky, Done, \& Pounds 1993). For this reason, we avoid
adjusting the spectral shapes at high energies to reproduce the detailed
shape of the CXB spectrum.

\subsection{Source counts}
\begin{figure}
\epsfysize=6.5in \epsffile{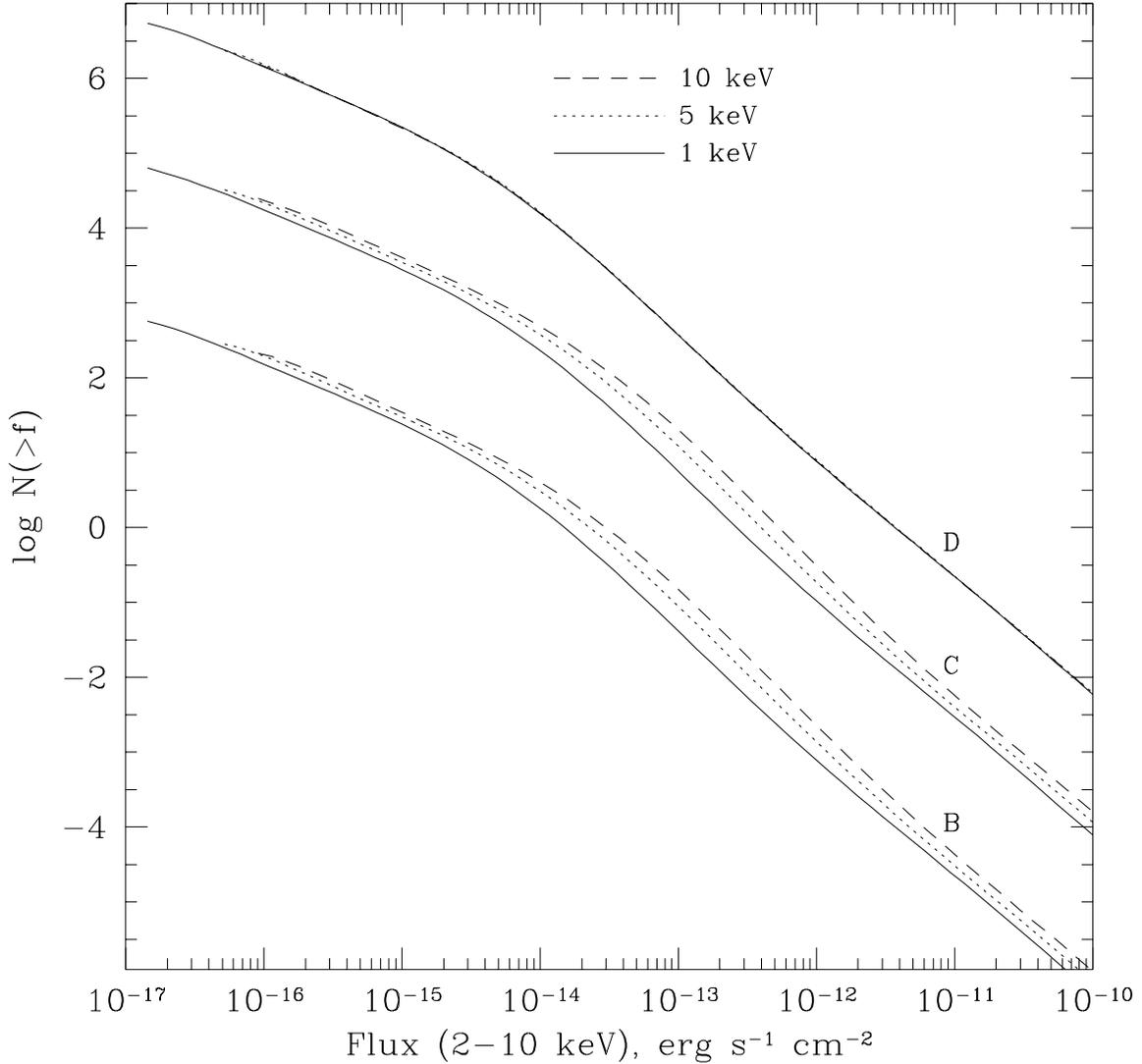}
\caption{
Source number-flux relation at three energies, 1, 5, and 10 keV, calculated
for different models of source spectra distribution. Normalizations are
arbitrary. For models with scatter of spectral parameters ({\bf B} and {\bf
C}) hard x-ray source counts yield more than two times more sources than the
observations performed at 1 keV; in the EMSS flux range, $10^{-13}-10^{-12}$
\ergs, the difference is approximately a factor of three for both models
with and without the soft excess ({\bf B} and {\bf C}, respectively).
Model {\bf D}, with the soft excess and without any scattering of spectral
parameters, predicts perfect agreement between soft and hard x-ray source
counts.}

\label{fig-counts}
\end{figure}

We calculated the source number-flux relations at three different energies,
1, 5, and 10 keV, by Monte-Carlo integration of eq.\ref{eq-counts}. Source
fluxes at these energies were transformed to 2-10 keV fluxes by means of
eq.\ref{eq-fluxtransform}, using the average source spectrum (i.e. the
spectrum defined by mean values of parameters in the appropriate model).
The results are presented in Fig. \ref{fig-counts}. Models {\bf A} and {\bf
B}, which are distinguished by Gaussian distribution of the soft excess
parameters, yield essentially the same results, and therefore, only model
{\bf B} is presented in Fig. \ref{fig-counts}.

Models with broad distributions of spectral parameters predict a significant
difference between the results of hard and soft x-ray source counts. In the
EMSS flux range, $10^{-13}-10^{-12}$ \ergs, there are approximately three
times more sources observed from the same population at 10 keV than at the
corresponding 1 keV flux. We emphasize that this effect does not arise from
the presence of any spectral features (like a soft excess). Model {\bf C},
having a simple power law spectrum for individual sources, predicts almost
the same difference in hard and soft x-ray source counts, as does model {\bf
B}, with the soft excess. As can be expected, model {\bf D} predicts
number-flux relations to be the same at all energies. The agreement is
partially distorted if the soft spectral component which we assumed to be
present below 1 keV starts at higher energies. In this case, the soft excess
in spectra of faint sources is shifted to low energies. As a result, when
using the $z=0$ average spectrum for flux transformations, the 2-10 keV flux
is underestimated, which causes an artificial decrease in the soft x-ray
source counts relative to hard x-ray counts. Calculations show however, that
in the EMSS flux range, $10^{-13}-10^{-12}$ \ergs, the difference is less
than a factor of 1.3 if a soft excess starts at energies of about 3 keV.

% % The effect of a soft excess is to enhance still further the difference at
% % low fluxes.  This is due to the redshift of the soft excess to very low
% % energies in the faintest sources, which are typically at higher redshifts.
% % This is well illustrated by model {\bf D} curves in Fig. \ref{fig-counts}.
% % In this model there is no distribution of spectral parameters, and 5 and 10
% % keV number-flux relations are in perfect agreement. The 1 keV \lognlogs~
% % curve decreases sufficiently only at fluxes less than about $10^{-13}$
% % \ergs. At this flux, many sources are at high redshift, and in the
% % observer's rest frame a sufficient part of the soft excess is shifted to
% % lower energies. As a result, when using the $z=0$ average spectrum for flux
% % transformations, the 2-10 keV flux is underestimated, which causes an
% % artificial decrease in the soft x-ray source counts relative to hard x-ray
% % counts.  However, in the EMSS flux range, $10^{-13}-10^{-12}$ \ergs, the
% % presence of a soft excess results in less than a factor of 1.3 difference.

\section {Discussion} \label{sec-discussion}

In this paper we have shown the importance of the distribution of spectral
parameters in calculations of the contribution of any source population to
the x-ray background. Of known classes of potential contributors to the CXB,
quasars demonstrate probably the broadest distribution of spectral
parameters, with 2-10 keV power law indices ranging from $\sim 0$ to $\sim
1.5-2$. Using models of QSO spectra distributed closely to what is observed,
we have shown that the composite spectrum of faint QSOs can be significantly
harder than that given by the average value, $\sim 0.8$, of the spectral
index. In fact, the composite spectrum of the QSO population closely mimics
the 0.5-100 keV CXB spectrum, if one assumes the existence of a high energy
cutoff in QSO spectra at around 100 keV. Another important effect which is
readily reproduced, when source spectra span a broad range, is the
difference between the results of hard and soft x-ray source counts.  The
same population of sources yields about three times more sources at around
10 keV than at the corresponding 1 keV flux.

The key point of our spectral distribution models is that the soft x-ray
luminosity, at around 1 keV, does not correlate with other spectral
properties of sources. This means that hard sources are brighter at higher
energies than soft sources, and this actually accounts for approximately
half the effect of hardening required to reproduce the CXB spectrum;
redshift effects serve to enhance the effect.  As was discussed in Section
\ref {sec-models}, available x-ray observations can neither prove nor
disaprove this assumption, and this is obviously the weak point of the
described model. Although the existence of a characteristic energy at which
quasar luminosities and spectra are not correlated does not seem
unreasonable, its particular value of 1 keV was chosen partially for the
sake of simplicity and partially because of the wish to reproduce the CXB
spectrum as well as possible. At the time of this writing it was generally
accepted that the CXB intensity in the soft x-ray band is in excess over the
$E^{-0.4}$ extrapolation from higher energies.  Note that models {\bf A}
through {\bf C} spectra show $\sim 40-50$\% excess over the $E^{-0.4}$
spectrum in the $1-2$ keV energy band, which is in agreement with ROSAT
measurements of the CXB spectrum (Hasinger 1992). Later ASCA measurements
(Gendreau et al. 1994), on the contrary, show that CXB follows the
$E^{-0.4}$ spectrum down to at least 1 keV. Our model still can fit this
spectrum (without the ``soft excess''), provided that quasar luminosities
are uncorrelated with spectra at lower energies, $0.1-0.3$ keV, as is
suggested by Laor et al. (1994).

% With the existing x-ray data
% it is difficult to verify our assumptions. In the soft x-rays, below 3 keV,
% where most quasars are observed, there is no obvious correlation between
% x-ray luminosity and spectral properties (Wilkes \& Elvis 1987).  However,
% the soft excess may comprise a significant part of the source intensity in
% this band, therefore it is not straightforward to extend the observed lack
% of correlation between luminosities and spectra to harder x-ray bands. In
% the standard x-ray band, 2-10 keV, quasars do demonstrate a negative
% correlation between spectral indices and hard x-ray luminosities (Williams
% et al. 1992). Unfortunately the Ginga sample of quasars consists of only
% about a dozen sources. New and more sensitive observations in the energy
% band up to at least 10 keV are required to refine models of QSO spectral
% distributions, and these can then be used to recalculate the contribution
% of quasars to the hard x-ray background.

% Spectral properties of quasars at energies higher than $\sim 20$ keV are
% very poorly known. Only a few of the most luminous sources have been
% observed so far.  Of course the average standard x-ray spectral index of 0.8
% requires that QSO spectra must eventualy steepen, but at present one can
% only speculate where the characteristic break energy lies.

This paper assumes that the high energy break in QSO spectra occurs at
around 100 keV. With this value of the break energy it is possible to
reproduce the CXB spectrum in the very broad energy range from 0.5 to about
100 keV, if the cosmological density parameter, $\Omega$ is zero. To
reproduce the CXB spectrum in $\Omega=1$ Universe a break energy of about 70
keV is required, in good agreement with the break energy of 50-100 keV
observed by OSSE in low luminosity AGNs, mostly Seyfert 1s (Johnson et al.
1994).

Summarizing, the paper shows that quasars still may be considered as the
main contributors to the broad-band x-ray background. It also shows the
importance of relaxing the assumption that all source have the same
spectrum when the source population is characterized by significant
distribution of spectral shapes.

%\clearpage
\begin {acknowledgements}
The author would like to thank R.Sunyaev for his friendly criticism. Special
thanks to W.Forman for his great help in work on the manuscript.
\end {acknowledgements}

\section*{References}
\beginrefs

Boyle B.J., Griffiths, R.E., Shanks, T., Steward G.C., and Georgantopoulous,
I. 1993, MNRAS, 260, 49.

Comastri, A., Setti, G., Zamorani, G., Elvis, M., Giommi, P., Wilkes, J., \&
McDowell, J., 1992, ApJ, 384, 62.

David, L.P., Slyz, A., Jones, C., Forman, W., Vrtilek, S.D., Arnaud, K.A.,
1993, ApJ, 412, 479

Fabian, A.C., Canizares, C.R., \& Barcons, X. 1989, MNRAS, 239, 15.

Fiore, F., Elvis, M., Siemiginowska, A., Wilkes, B.J. \& McDowell, J.C. 1994
ApJ in press.

Gendreau, K., et al. 1994, ``New Horizons of X-ray Astronomy -- First
Results from ASCA'', preprint.

Giacconi, R., Gursky, H., Paolini, F., and Rossi, B. 1962, Phys. Rev.
Letters, 9, 439.

Giacconi, R., \& Zamorani, G. 1987, ApJ, 313, 20.

Gioia, I.M., Henry, J.P., Maccacaro, T., Morris, S.L., \& Stocke, J.T. 1990,
ApJ, 356, L35.

Hasinger, G. 1992, in: The X-ray Background, X. Barcons \& A.C. Fabian eds.
(Cambridge University Press), p.1.

Hasinger, G., Burg, R., Giacconi, R., Hartner, G., Schmidt, M., Tr\"umper,
J., \& Zamorani, G., 1993, A\&A, 275, 1.

Johnson, W.N. et al., 1994, in the AIP Conference Proceedings 304, eds. C.
Fitchel, N. Gehrels, J. Norris, 515.

Jourdain, E., et al. 1992, A\&A, 256, L38.

Laor, A., Fiore, F., Elvis, M., Wilkes, B., \& McDowell, J.C., 1994, ApJ,
435, 611.

Marshall, F. et al. 1980. ApJ, 235, 4.

Mather, J. et al. 1990, ApJ, 354, L37.

Matsuoka, M., Piro, L., Yamauchi, M., and Murakami, T. 1990, ApJ,
361, 440.

Mattig, W. 1958, Astron.Nachr. 284, 109.

Mushotzky, R.F., 1984, Adv. Space. Res., 3, 10.

Mushotzky, R.F., Done, C., \& Pounds, K. 1993, Ann. Rev. Astr. \& Ap., 31,
717.

Nandra, K., \& Pounds, K.A., 1994, MNRAS, 1994, 268, 405.

Piccionotti, G., Mushotzky, R.F., Boldt, E.A., Holt, S.S., Marshall, F.E.,
Serlemitsos, P.J., and Shafer, R.A. 1982, ApJ, 253, 485.

Piro, L., Matsuoka, M., and Yamauchi, M. 1989, Proc. 23rd ESLAB
Symposium on ``Two Topics in X-ray Astronomy'', ESA SP-296, 819.

Pounds, K., Nandra, K., Stewart, G., George, I., and Fabian A. 1990,
Nature, 344, 132.

Schwartz, D. and Tucker, W. 1988, ApJ, 332, 157.

Wilkes, B. and Elvis, M. 1987, ApJ, 323, 243.

Williams, O.R. et al., 1992, ApJ, 389, 157.

\endrefs
\clearpage

\end{document}